\documentclass[aps,preprint]{revtex4-1}
\usepackage{graphicx}  
\usepackage{amsmath} 
\usepackage{dcolumn}   
\usepackage{bm}        
\usepackage{amssymb}   
\usepackage{epstopdf}
\newcommand{\bib}{\bibitem}
\def\vF{v_F}
\begin{document}

\title{Granular topological insulators}
\author{Abhishek Banerjee$^1$, Oindrila Deb$^2$, Kunjalata Majhi$^1$, 
R. Ganesan$^1$, Diptiman Sen$^2$, and P. S. Anil Kumar$^{1,3}$}
\affiliation{\small{$^1$Department of Physics, Indian Institute of Science, 
Bengaluru 560 012, India \\
$^2$Centre for High Energy Physics, Indian Institute of Science, Bengaluru
560 012, India
\\ $^3$ Centre for Nano Science and Engineering, Indian Institute of Science, Bengaluru
560 012, India} }

\date{\today}

\begin{abstract}

Granular conductors form an artificially engineered 
class of solid state materials wherein the microstructure 
can be tuned to mimic a wide 
range of otherwise inaccessible physical systems. At the same time, 
topological insulators (TIs) have become a cornerstone of modern 
condensed matter physics as materials hosting metallic states on the 
surface and insulating in the bulk. However it remains to be understood 
how granularity affects this new and exotic phase of matter. We perform 
electrical transport experiments on highly granular topological insulator 
thin films of Bi$_2$Se$_3$ and reveal remarkable properties. We observe 
clear signatures of topological surface states 
despite granularity with distinctly different properties 
from conventional bulk TI systems including sharp surface state 
coupling-decoupling transitions, large surface state penetration depths 
and exotic Berry phase effects. We present a model which explains these 
results. Our findings illustrate that granularity can be used 
to engineer designer TIs, at the same time allowing easy 
access to the Dirac-fermion physics that is 
inaccessible in single crystal systems.

\end{abstract}

\maketitle


In solid state systems, granularity can be conceptually introduced by 
considering nanoscale sized grains that are packed to form a 
macroscopic system. While the constituent granules are large 
enough to retain the characteristic properties of the parent material, 
they are also subject to strong quantum confinement. Depending on how 
the granules couple to their neighbours, a host of physical systems can be 
realized~\cite{grreview, gr1, gr2}. Granular media have therefore been used 
as a test bed for studying a rich variety of physical phenomena including
metal-insulator transitions~\cite{gr3,gr3_1}, superconductivity~\cite{gr4} and magnetism~\cite{gr5}, apart from a versatile range of applications including solar cells, light emitting diodes, photodetectors and field-effect transistors~\cite{app1, app2, app3}.
 
The extension of these ideas to the case of three-dimensional topological 
insulators (TIs) however remains unknown. In the recent surge of interest 
in this novel phase of matter~\cite{tireview1,tireview2}, a lot of 
experimental~\cite{chen09,zhang09a,xia09,hsieh09a,roushan09,hsieh09b,peng10} 
and theoretical~\cite{fu07,kane07,roy09,teo08,qi08,zhang09b,liu10}
activity has been directed at understanding their physical properties,
the most interesting of which is the presence of gapless
Dirac fermions on the surfaces of these materials. The chiral nature of 
these surface fermions follows from the strong spin-orbit coupling and
ensures that the momentum of an electron fixes the direction of its spin.
The situation is however not so simple when one deviates from an ideal TI. 
For example, it has been shown that bulk disorder may completely randomize 
the surface spin polarization~\cite{soriano}. Also, strong bulk-to-surface 
coupling in bulk-doped TIs may delocalize surface 
states~\cite{bulk_surf_hybridization}. 
It is not clear whether such surface states can manifest in 
highly granular TIs. This motivates us to ask the following questions: (i) 
how does granularity affect TIs, and (ii) can granularity be used to tune 
the properties of TIs? We will concentrate on the specific case of 3D TIs
in this work.

\begin{figure*}[!t]
\includegraphics[width=1.0\linewidth]{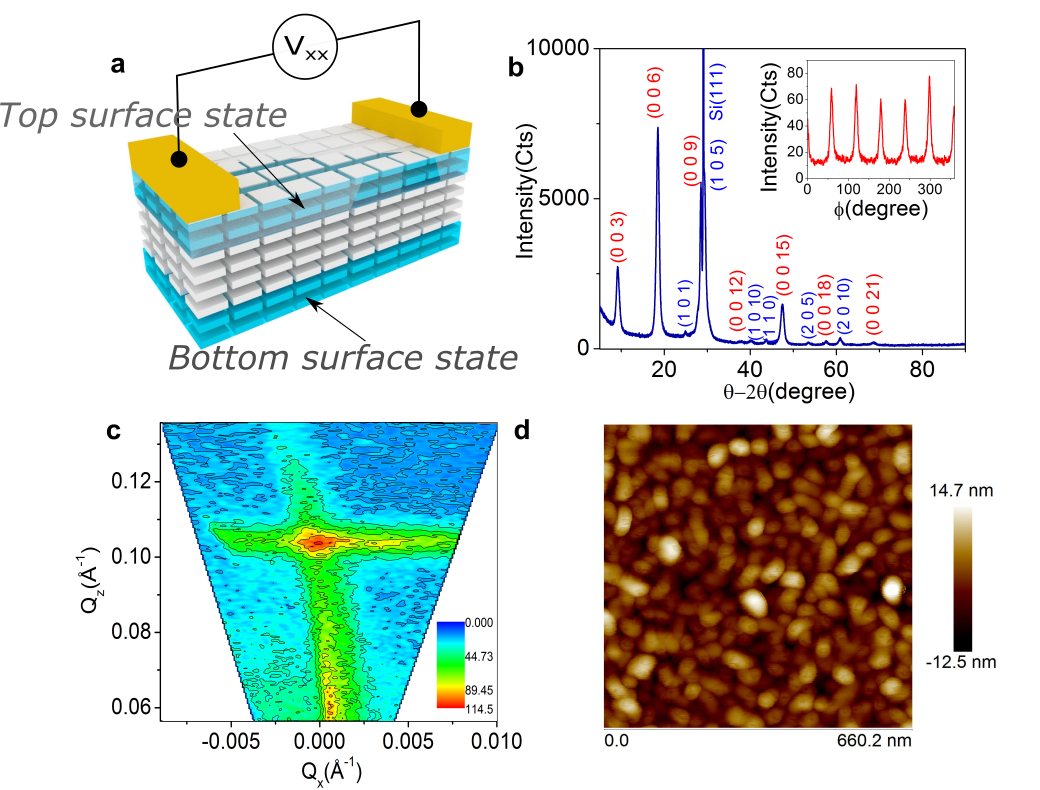}
\caption{ {\bf Growth of granular TI thin films} (a) Schematic of a granular conductor composed of ordered TI 
granules. Under appropriate conditions, the entire conductor behaves as a 
single crystal TI with distinct top and bottom surface states. The finite 
penetration depth of the surface states is shown in blue. (b) Standard 
$\theta-2\theta$ XRD scan of a typical granular thin film shows that (003n) peaks (marked in red) are dominant implying a preferential orientation in the $c$-axis direction. Non-oriented peaks (marked in blue) reflect the granularity of our samples. Inset: Azimuthal scan of (105) planes of Bi$_2$Se$_3$ 
shows six-fold symmetry and a large FWHM $\sim$ 8$^\circ$. (c) Reciprocal space maps of (003) peak peak indicate highly granular thin films. The large broadening of the reciprocal lattice peak is indicative of a highly granular material, with 
grain sizes that roughly scale as the inverse of the peak broadening. (d) AFM image showing the surface topography of a granular sample.} 
\label{fig01} \end{figure*}


Granular Bi$_2$Se$_3$ thin films are grown on high resistivity Si(111) 
substrates using pulsed laser deposition. The degree of granularity is controlled by tuning the laser pulse energy and is
quantified using x-ray diffraction and atomic force microscopy (see Supplementary Information for details).
Our measurements 
indicate an in-plane coherence length $\xi_\parallel \approx 10$ nm, which is 
the same as the in-plane grain size.The vertical grain size is smaller 
and is estimated to be $\sim$ 2-3 nm.
Our thin films are therefore composed of highly ordered aggregates 
of granules and can be schemicatically depicted as shown in Fig. 1 (a). 
Although the granular packing is largely uniform, we expect an asymmetry in 
the inter-granular coupling. While the in-plane coupling is expected to be 
strong due to a large density of lateral dangling bonds, the absence of 
dangling bonds in the vertical directions makes the vertical inter-grain 
coupling weak.

\begin{figure}[!t]
\includegraphics[width=1.0\linewidth]{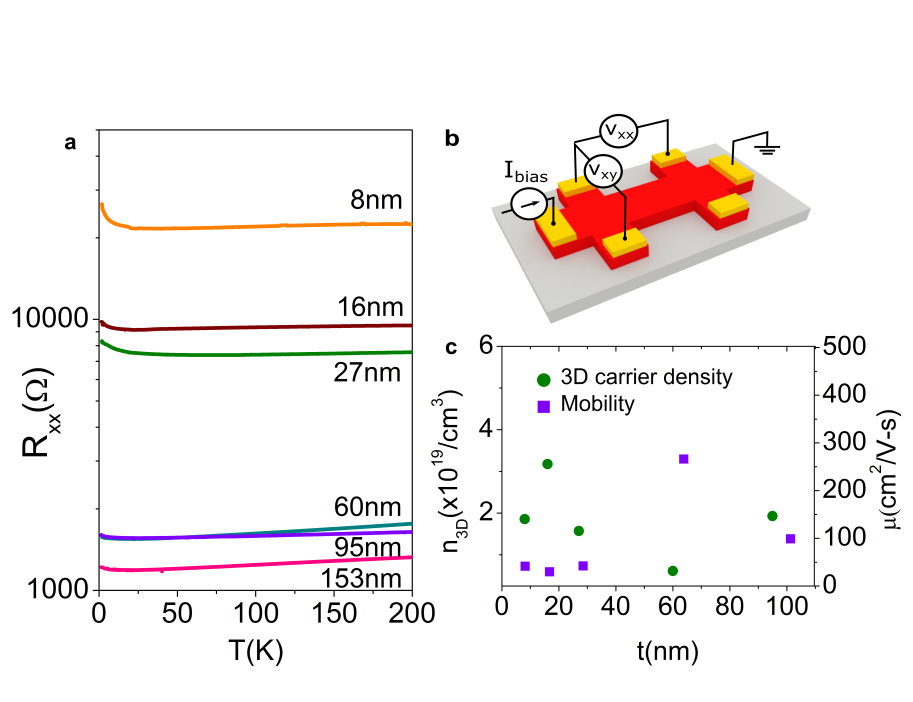}
\caption{{\bf Hall conductivity and resistivity of granular TI conductors} (a) Resistance versus temperature curves for granular samples with 
different thicknesses. (b) Schematic of the six probe Hall bar configuration 
that is used to measure the longitudinal and transverse resistance 
independently. (c) Hall carrier density and mobility for different sample 
thicknesses at $T=2$ K.} \label{fig02} \end{figure}


Electronic transport measurements 
are carried out on granular TI films with thickness ranging from 8 nm to 
153 nm. The resistance versus temperature curves for granular TI samples 
shown in Fig. 2 (a) suggest weakly metallic conductivity for all samples. 
At low temperatures a strong insulating upturn is seen below 
$\sim$ 30 K. Both the resistance minimum and the onset of the
upturn are enhanced for films with lower thickness. This is a result of 
inter-grain hopping (discussed later) of electrons that is the dominant 
transport mechanism at low temperatures~\cite{grreview}. At higher temperatures, however, 
the conductance is determined instead by the intra-grain conductance that is 
limited by long-range impurity scattering from charged defect centres. 
A striking feature of our data is that for films with higher thickness
(60 nm - 153 nm), the conductance does not vary much with the sample thickness.
This is indicative of surface confined transport in the thick film regime 
where the top and bottom surface states do not overlap. 
This is surprising given that the multiply connected geometry 
of our thin films is naively not expected to give rise to well-defined 
surface excitations. 

Hall effect measurements (at 2 K) are used to estimate the disorder introduced 
by granularity. The values of carrier density ($n_{3D}$)
and mobility ($\mu$) are shown in Fig. 2 (c). For a typical sample, assuming 
a spherical Fermi surface we estimate a Fermi momentum 
$k_F=(3 \pi^2 n_{3D})^{1/3} \sim 0.8-0.9$ ${\rm nm}^{-1}$ and a
mean free path $l_e=(\hbar \mu/e) k_F \sim 2-3$ nm.
The extent of disorder in our 
samples can be estimated by using the Ioffe-Regel criterion for which 
$k_F l_e\gg1$ implies weakly disordered conduction while $k_F l_e \ll 1$ 
implies strong localization. For our samples, $k_F l_e \sim 2-3$ which puts 
our samples in a regime proximate to the localization transition. The small 
grain size limits the mean free path of electrons. While intra-grain transport 
remains diffusive, the proximity to localization is a consequence of limited 
inter-grain transport which proceeds through electron tunneling across 
barriers. 
At low temperatures the tunneling is 
suppressed giving rise to an insulating upturn in resistivity. While the 
effects of granularity are already evident at this level, the most 
surprising effects are revealed by magnetoresistance measurements.

\begin{figure*}[!t]
\includegraphics[width=1.0\linewidth]{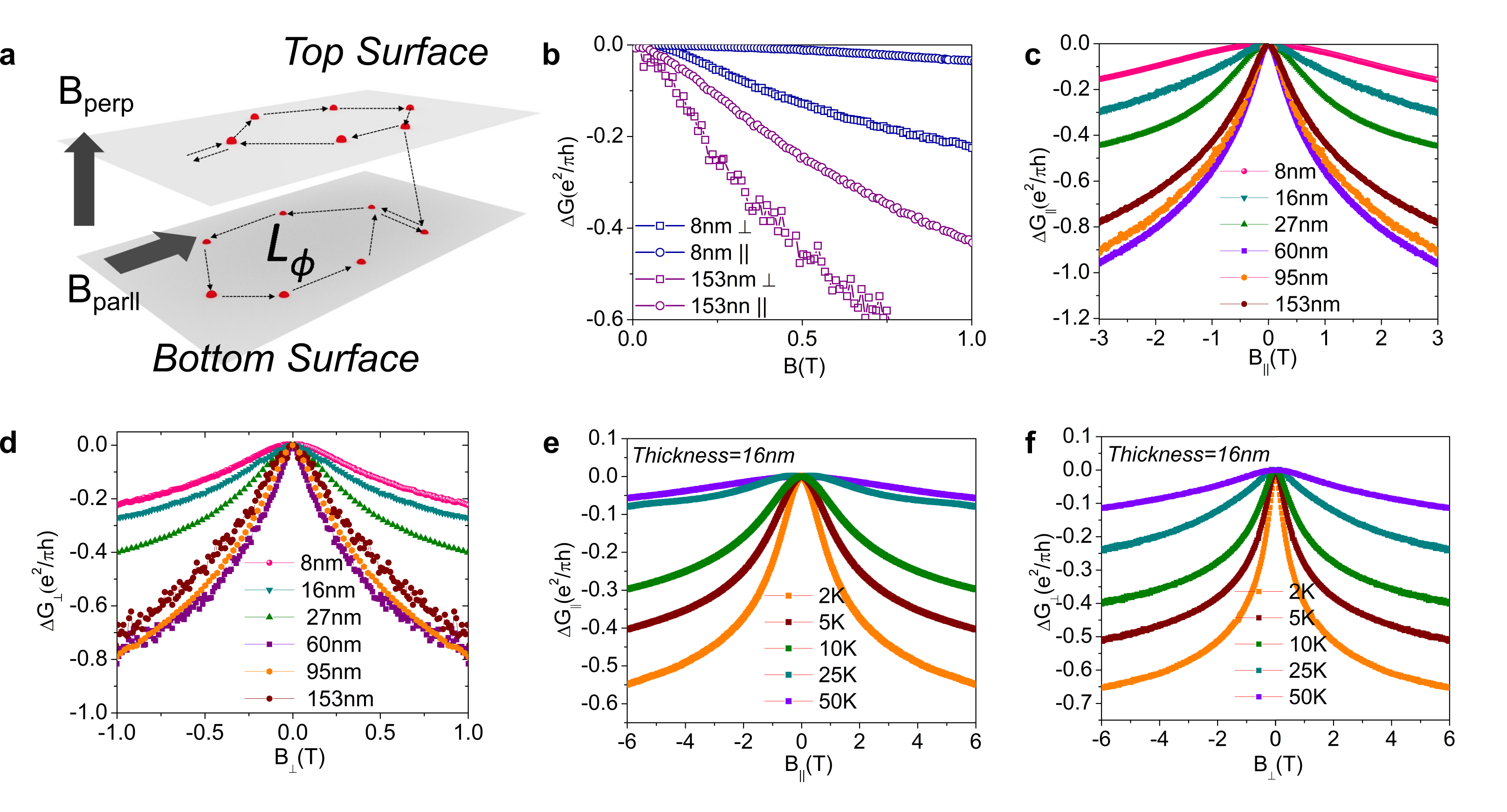}
\caption{{\bf Weak Anti-localization correction to conductance} (a) Mechanism of WAL due to suppressed back-scattering. (b) 
Low-field magnetoconductance for $t=8$ nm and $t=153$ nm films 
for in-plane and out-of-plane field configurations. The magnetoconductance 
increases with thickness. Also, the low-field in-plane magnetoconductance is 
smaller in magnitude than the out-of-plane magnetoconductance. Thickness 
dependence of magnetoconductance for (c) parallel and (d) perpendicular 
fields. Temperature 
dependence of high-field magnetoconductance for (e) parallel and (f) 
perpendicular fields for a $t=16$ nm film.} \label{fig03} \end{figure*}

A strong manifestation of the 
topological properties of surface states obeying the Dirac equation 
is the Berry phase of $\pi$ associated with them~\cite{fu07,xia09}. This is 
a consequence of the helical spin structure 
of the electrons in momentum space. Such an electron traversing a path 
which encircles the origin in momentum space, accumulates a phase of $\pi$ 
upon completing the loop. This leads to a suppression of coherent 
back-scattering (Fig. 3 (a)). 
By applying a magnetic field, time reversal symmetry is broken 
and WAL can be destroyed giving rise to a negative magnetoconductance. 
All our samples indicate a sharp dip in the conductance with increasing 
magnetic field, indicating WAL (Figs. 3 (b-f)). For magnetic fields 
perpendicular to the sample, the magnitude of the dip increases with 
increasing thickness in the thin limit, but almost saturates for the thicker 
films (Fig. 3 (d)). In-plane field measurements however yield a surprisingly 
large magnetoresistance which is a consequence of predominant surface state 
transport as discussed later. We perform separate 
analysis for perpendicular field and parallel field magnetoconductance data (see Supplementary Information for details). 
The perpendicular field magnetoconductance data are fitted to the 
Hikami-Larkin-Nagaoka (HLN) equation that quantifies the conductance 
correction due to WAL in two dimensions~\cite{hikami80,altshuler80,garate12},
\begin{equation} 
\Delta \sigma_\perp (B_\perp)=\alpha_\perp \frac{e^2}{2\pi^2 \hbar} \left[
\psi\left(\frac{1}{2} + \frac{\hbar}{4el_\phi^2B_\perp}\right)-\ln \left( 
\frac{\hbar}{4el_\phi^2B_\perp}\right)\right] \end{equation}
in the limit of strong spin-orbit scattering, where $\sigma$ is the sample 
conductivity, $B_\perp$ is the applied magnetic field, 
$l_\phi$ is the phase coherence length, and 
$\psi(x)$ denotes the digamma function. While the perpendicular field 
magnetotransport gives information about the number of phase 
coherent channels and the associated decoherence lengths, the parallel field 
magnetotransport is a powerful tool for studying the spatial 
extent of the channels. In conventional 
TIs, the surface state penetration depth is small and therefore a parallel 
magnetic field does not decohere these channels. The only conductance 
correction therefore comes from the bulk of the sample. The complicated 
coupling between bulk and surface transport has however made previous 
experimental findings ambiguous, and it has been pointed out that the parallel 
and perpendicular field magnetotransport cannot be consistently 
explained~\cite{parallel_field_MR}. On the contrary, we find that conductance 
corrections arising out of bulk electrons in our samples are negligible, and 
clean signatures of surface state transport are obtained even in parallel 
field magnetoresistance measurements. We fit our parallel field data to the 
corresponding WAL equation for in-plane fields~\cite{parallel_field_fit},
\begin{equation}
\Delta \sigma_\parallel (B_\parallel)=\alpha_\parallel \frac{e^2}{2\pi^2 \hbar} \ln
\left(1+B^2_\parallel/B^2_\phi \right), \end{equation}
where $B_\phi = \hbar/(e \lambda l_\phi)$, $\lambda$ being the surface state 
penetration depth. In fitting our data, we take $\alpha_\parallel$ and 
$\lambda$ as fitting parameters while the phase coherence length $l_\phi$ 
is used from the perpendicular field data. We assume that the phase coherence 
length for the two measurements must be same, if the contributions to 
conductance corrections come from the same channels in both cases. 

\begin{figure}[!t]
\includegraphics[width=1.0\linewidth]{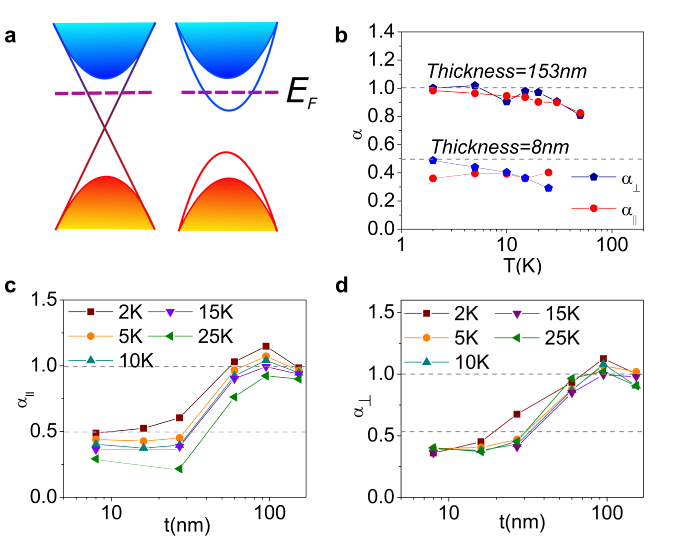}
\caption{{\bf Transition from coupled to decoupled surface states} (a) Coupling of surface states causes hybridization between the two surface states and opens up a gap at the Dirac point. (b) $\alpha$ is shown as a function of temperature for 153 nm film in the decoupled regime and 8 nm 
film in the coupled regime. Both out-of-plane and in-plane field measurements give similar values of $\alpha$ within experimental accuracy. As a function of increasing 
thickness, (c) $\alpha_\parallel$ and (d) $\alpha_\perp$ undergo a cross-over 
from 0.5 indicating coupled surface states to 1.0 indicating decoupled surface 
states, with a decoupling transition at $t\simeq 60$ nm.} \label{fig04} 
\end{figure}
 
The result of our analysis is shown in Fig. 4. In the thick-film regime 
(Fig. 4 (b)), $\alpha_{\perp/\parallel} \simeq 1$ for all ranges of 
temperature. More 
strikingly, the match between $\alpha_\perp$ and $\alpha_\parallel$ suggests 
that in both field configurations, the same channels contribute to WAL. 
Additionally $\alpha_{\perp/\parallel} \simeq 1$ indicates that both parallel 
and perpendicular magnetic fields interact with two distinct transport 
channels indicating a pair of decoupled topological surface states at
the top and bottom of the film. In the 
thin film regime, perpendicular and in-plane field measurements 
again give the same values of $\alpha$, but with $\alpha_{\perp/\parallel}
\simeq 0.5$, indicating a coupling of the two surface states.
Measurements over an entire thickness range indicate that 
this cross-over happens at around $t \approx 60$ nm (Figs. 4 (c-d)). 
Such clear signatures of coupling-decoupling transitions, for both in-plane 
and out-of-plane field configurations, have not been observed before for single
TI thin films. Most transport studies on single TI films indicate a strong 
coupling between opposite surfaces even for thick films where direct tunneling 
due to overlap of surface state wave functions is not possible. 
Additionally, no signatures of the surface state transport 
have been observed in in-plane field magnetotransport. Our samples on the other
hand show strong indications of surface state dominated transport, in both 
in-plane and out-of-plane measurement configurations.

The decoupling transition observed at $t \approx 60$ nm cannot be explained 
by existing ideas about TIs. We now show that the decoupling at such a large 
thickness is a consequence of the large penetration depth of surface states. 
As discussed before, the in-plane field magnetoresistance depends critically 
on the cross-section of the states with which it interacts. For surface 
states, this cross-section is proportional to the surface state penetration 
depth for thick films. In the thin film limit, where surface states exist 
throughout the volume of the film, the cross-section is proportional to 
the thickness of the film. Fig. 5 (a) shows the extracted parameter $\lambda$ 
which measures the extent of surface states in the $z$-direction. 
In the thick film limit and low temperature, $\lambda \approx 33$ nm. This is 
approximately half the sample thickness for which we observe a decoupling 
transition. This is expected: the coupling transition commences when the 
thickness of the film is such that the top and bottom surface states extend 
throughout the bulk of the sample. This large penetration depth allows the 
surface states to interact even with a parallel magnetic field and give an 
unexpectedly large parallel field magnetoresistance. This allows us not only 
to accurately predict the surface state penetration depths, but also study 
their temperature dependence. The temperature dependence of $\lambda$ shows 
a remarkable decrease with temperature for the thick films. On the other 
hand, for our thin samples $\lambda$ is almost equal to the sample thickness 
and is very weakly dependent on temperature, most prominently observed for 
the thinnest measured film at $t=8$ nm. The unusually large penetration 
depths and their temperature dependence is a consequence of 
the granularity as described below.

\begin{figure}[!t]
\includegraphics[width=1.2\linewidth]{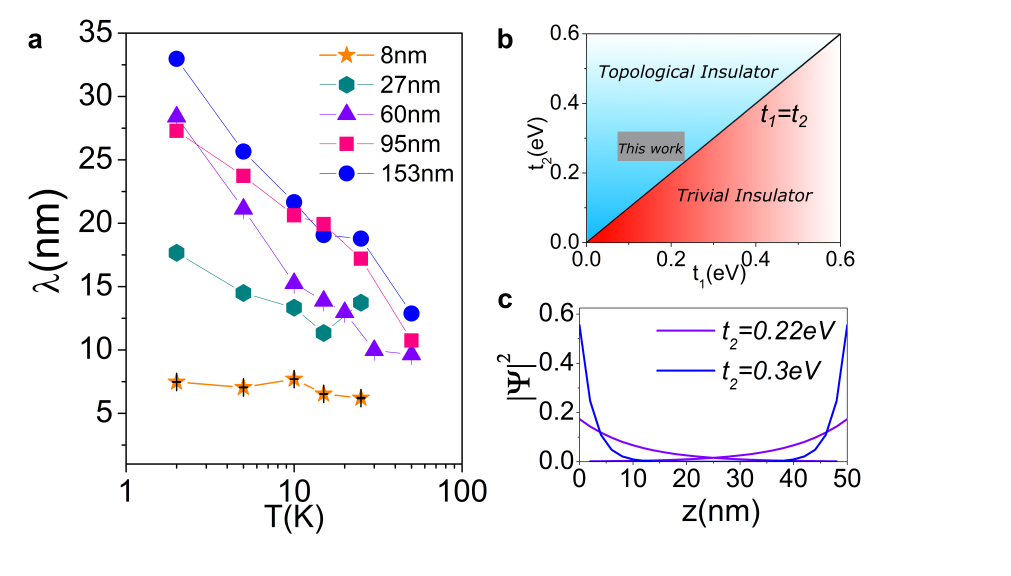}
\caption{{\bf Surface state penetration depth in the TI regime} (a) Surface state penetration depth $\lambda$ for 
different sample thicknesses and temperatures. For samples in the thick-film 
limit, $\lambda$ decreases with increasing temperature from $\sim 30$ nm to 
$\sim 15$ nm. For thin samples, the temperature dependence of $\lambda$ is 
weak and approaches the sample thickness. This allows us to estimate the 
ratio of intra-grain coupling ($t_1$) to inter-grain coupling ($t_2$). 
(b) shows that the interplay between $t_1$ and $t_2$ can give rise to a 
trivial insulator phase when $t_1 > t_2$ and topological insulator phase 
when $t_2 > t_1$. Our samples lie in the gray shaded area of this phase 
diagram. The ratio of $t_1$ to $t_2$ drastically modifies the surface 
state penetration depth. (c) shows the penetration of the $z$-component of 
top and bottom surface state wave functions into the bulk of the TI for two 
different values of $t_2$ at a constant value of $t_1=0.2$ eV.} 
\label{fig05} \end{figure}


We can understand the various experimental results using
a simple model of a granular TI (see Supplementary Information for details). Guided by
the experimental observations that the grains are quite well-ordered and
that the in-plane grain size is much larger than the perpendicular thickness,
we consider a film consisting of a regular array of grains which are 
infinitely large in-plane
(called the $x-y$ surface) and repeat periodically in the $z$-direction 
(along the $c$-axis) with a unit cell size $L \sim 2$ nm. We 
introduce intra-grain couplings $t_1$ and 
inter-grain couplings $t_2$. ($t_1$ will depend 
strongly on the individual grain sizes; increasing the size of a grain 
increases the distance between its opposite surfaces and therefore decreases 
$t_1$). We consider surface states with a momentum ${\vec k} = (k_x, k_y)$.
If the number of grains $N \gg 1$ (a thick film), we can introduce a Bloch 
momentum $k_z$ in the $z$-direction (lying in the range $[-\pi/L, 
\pi/L]$). We then find a continuum of states with $E_{{\vec k}, \pm} = \pm 
\sqrt{ \hbar^2 \vF^2 {\vec k}^2 + t_1^2 + t_2^2 + 2 t_1 t_2 \cos (k_z L)}$; 
this has a gap given by $2 |t_1 - t_2|$. (Here $\vF$ is the Fermi velocity on 
the $x-y$ surface and is given by $0.333$ eV-nm$/\hbar$ for 
Bi$_2$Se$_3$~\cite{tireview2}). These form the bulk states of the film. 

Next, if $t_1 < t_2$, we find states which are localized near the top and 
bottom surfaces of the film; these have the dispersion $E_{{\vec k},\pm} = \pm
\hbar \vF |{\vec k}|$ and their wave functions decay into the bulk of the film
with a penetration depth $\lambda$ given by $e^{-L/\lambda} = t_1 /t_2$.
These are the surface states of the film (Fig. 5 (c)). There are no such states
if $t_2 < t_1$. Thus the entire film behaves like a single TI with gapped 
bulk states and gapless surface states, and there is a quantum phase 
transition from a non-topological phase at $t_2 < t_1$ to a topological
phase at $t_1 < t_2$. We now note that since the inter-grain coupling $t_2$ 
arises due to tunneling through a barrier given by the charging energies
of two grains, it increases with the temperature in an activated form. 
On the other hand, the intra-grain coupling $t_1$ does not change the 
charging energy and is not expected to vary significantly with the temperature.
Hence we expect $t_1/t_2$ and therefore $\lambda$ to decrease with increasing
temperature which agrees with our observations. We observe from Fig. 5 (a) that 
for the thickest film, $\lambda$ varies from about 33 nm at 2 K to 12 nm at 
50 K. Taking the vertical grain size to be about 2 nm, the expression
$e^{-L/\lambda} = t_1 /t_2$ implies that $t_1 /t_2$ varies from $0.85$
to $0.94$; hence the change with temperature is not large.

Finally, if $t_1 < t_2$ and the number of grains $N$ is not very large, then 
there will be a significant hybridization between the states at the top and 
bottom surfaces. One can show that this is given by $\gamma \sim (t_1/t_2)^N$;
the surface states then have the dispersion $E_{{\vec k},\pm} = 
\pm \sqrt{\hbar^2 \vF^2 {\vec k}^2 + \gamma^2}$ which has a gap $2 |\gamma|$.
We see that $\gamma$ decreases both with increasing temperature (since
$t_2$ increases) and increasing film thickness (since $N$ increases). 

In contrast, the surface states in conventional TIs are confined within 
$\sim$ 2-3 nm of the sample surfaces. 
The novelty of our TI is exemplified by the fact that the ratio of inter-grain
to intra-grain tunneling determines the surface state penetration depth.
When the two coupling strengths are nearly equal, the system is close to a 
topological phase transition and exhibits large surface state decay lengths. 
On increasing the temperature, the inter-grain coupling becomes stronger;
this gives rise to a smaller 
penetration depth as we showed above. For thinner films, the top 
and bottom surface states are strongly coupled and span the entire bulk 
of the film; $\lambda$ is then about the same as the film
thickness. The response to temperature is weak in this limit. 
This counter-intuitive observation suggests that increasing the temperature 
makes the system become more `topological'. While we show that 
temperature can be used as a control for `topology' in our system, it suggests 
that an ability to vary the ratio of the inter-grain coupling to the 
intra-grain coupling can be used to tune the various system parameters. 

\begin{figure}[!t]
\includegraphics[width=1.3\linewidth]{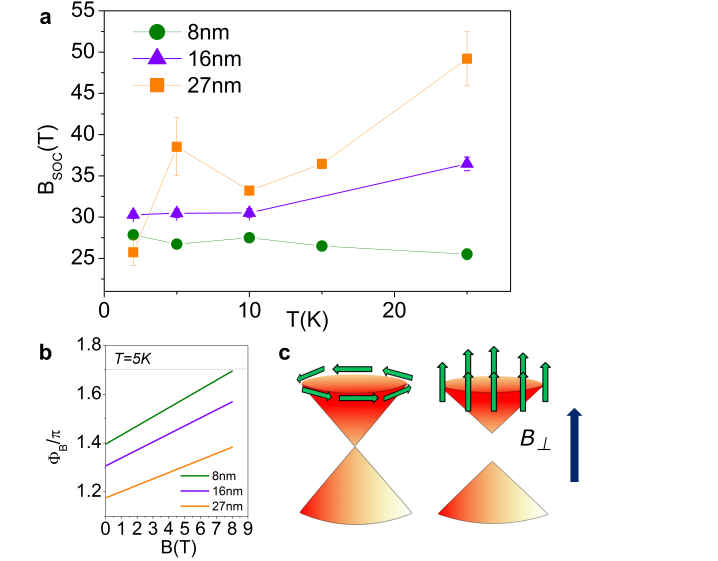}
\caption{{\bf Berry phase effects} (a) Effective spin-orbit field as a function of temperature in the 
low-thickness limit. (b) Extracted dependence of the surface state Berry 
phase on magnetic field and sample thickness at T=5K. We chose only one 
state with all signs positive in Eq. (3). (c) Schematic showing how an 
external magnetic field $B_\perp$ can compete with the intrinsic spin-orbit 
field to destroy the chiral spin momentum locking and give rise to a trivial 
Berry phase.} \label{fig06} \end{figure}

We now show how granularity can be used to vary the Berry phase.
This phase can get modified from the 
ideal value of $\pi$ for two reasons. First, the tunneling $\gamma$ 
between the surface states effectively adds a mass term to the Dirac 
Hamiltonian describing those states; this modifies the Berry phase.
The mass term also destroys ideal spin-momentum locking and effectively 
weakens the effect of the spin-orbit field $B_{SOC}$ that quantifies this 
locking. While ARPES and STM measurements indicate a loss of spin-momentum 
locking for ultra-thin films~\cite{ARPES,STM}, a quantification of this loss 
has been missing in earlier work. Secondly, the application of a magnetic 
field that is comparable in strength to $B_{SOC}$ causes the spins to get 
aligned in the direction of the magnetic field; this also modifies 
the Berry phase from $\pi$. Combining the effects of tunneling $\gamma$ 
between the top and bottom surface states and a Zeeman magnetic field $B_\perp$ 
applied in the $z$-direction, we find that there are four states $\Psi_j$
for a given momentum $k_F$. The Berry phases in these states are given by
\begin{equation} \Phi_j = \pi \left( 1 \pm \frac{b_z \pm \gamma}{\sqrt{ 
\hbar^2 \vF^2 k_F^2 + (b_z \pm \gamma)^2}} \right), \end{equation}
where $b_z = g \mu_B B_\perp/2$ ($g$ and $\mu_B$ are the gyromagnetic ratio
and Bohr magneton respectively). Experimentally, this manifests as a sharp 
degradation of WAL at higher fields. The standard HLN equation can be 
used to fit the experimental data at small fields; however for large fields 
a large deviation is seen between the HLN equation and our data, 
with the HLN equation predicting a higher magnetoconductance than is observed. 
This strong suppression of WAL observed in the thin film limit can be 
explained by introducing a correction to the HLN equation which depends
on the effective spin-orbit field $B_{SOC}$. Similar to prior works 
that have quantified the effect of Berry phase on WAL~\cite{WAL_HgTe, Shen_WAL}, we 
use a phenomenologically modified HLN equation (see Supplementary Information for details)
to capture this effect:
\begin{eqnarray}
\Delta \sigma_{\perp}(B_\perp) &=& \alpha_{\perp}\left[1-2 (B_\perp/B_{SOC})^2\right]
\frac{e^2}{2\pi^2 \hbar} \nonumber \\
&& \times \left[ \psi\left(\frac{1}{2} + \frac{\hbar}{4el_\phi^2B_\perp}\right) 
-\ln\left(\frac{\hbar}{4el_\phi^2B_\perp}\right) \right]. \end{eqnarray}
This allows us to extract the effective spin-orbit field $B_{SOC}$ 
(Fig. 6 (a)). We find a 
strong thickness dependence of $B_{SOC}$ with a value as small as $\sim 30$ T 
for $t=8$ nm at 2 K. Further, we observe that this increases rapidly with 
increasing temperature and thickness. This can be explained as follows. The 
Fermi energy of our system is fixed by the Fermi level which is pinned to 
the conduction band bottom of the bulk states which
come from the individual TI grains. The increase in mass
of the Dirac fermions with decreasing thickness leads to an upward movement 
of the massive Dirac bands with the Fermi energy kept fixed. This gives rise 
to a decrease in the value of the surface $k_F$ and a corresponding decrease 
in $B_{SOC} \propto \vF k_F$, accordingly modifying the Berry phase (Fig. 6 (b)). Such a competition of 
the Zeeman field with the internal spin-orbit field is difficult to observe 
in single crystal TIs because intrinsic SOC fields are of the order 
of several hundred Teslas. Introducing granularity decreases the effective 
spin-orbit field, and makes the study of interesting Dirac fermion physics 
experimentally feasible.


Granularity opens up an altogether different route to control the properties 
of TIs. We show that a multiply connected system composed of individual TI 
grains can, under certain conditions, behave as a single crystal TI but with
spectacularly different properties. Our granular thin films show distinct 
signatures of top and bottom surface states that are decoupled for thick films 
and get coupled below a critical film thickness. In stark contrast to single 
crystal TIs, these surface states exhibit a large penetration depth determined 
by grain size and inter-grain coupling and can be tuned by temperature;
remarkably,
the penetration depth is much larger than the thickness of a single grain. The 
Dirac nature of these states is clearly observed by studying the modification 
of their associated Berry phase in the presence of an external Zeeman field 
and tunneling between surface states, allowing a quantification of the 
intrinsic spin-orbit field. 
Future experimental and theoretical work can focus on careful tuning of the 
ratio of intra-grain coupling to inter-grain coupling that can drive such a 
system from a non-TI to a TI phase, allowing access to exotic topological 
phase transitions. Granularity may also be used to study topological Anderson 
insulators that are yet to be experimentally realized. Finally, the 
manipulation of spin transport in designer TIs may prove very interesting 
for the spintronics community.

\vspace{1cm}

{\bf Methods}

\vspace{0.5cm}

{\bf Thin film growth:} Thin films of Bi$_2$Se$_3$ were grown using pulsed 
laser deposition. The Bi-Se target was prepared by heating Bi (99.999\%) and 
Se (99.999\%) at 1123 K for 24 hours followed by slow cooling. Target 
composition was adjusted to compensate for Se loss during ablation. High 
resistivity (10 $\Omega$-cm) Si(111) substrates were used for deposition at 
a subsrate temperature of $\sim$ 523 K. The thin films were characterized 
using XRD, AFM, STM and Auger electron spectroscopy. We adjust the film growth 
rate by tuning the laser pulse energy, which also changes the morphology of 
the film from highly crystalline at low deposition rates to highly granular 
at high deposition rates. Granular films are grown at a high growth rate of 
$\sim$ 60 pulses/nm. The thickness of the films is modified by changing the 
number of pulses fired, and verified using AFM measurements.

\vspace{0.5cm}

{\bf Transport measurements:} Electrical transport measurements were carried 
out on six-probe Hall bars. The Hall bars were prepared by shadow deposition 
using metallic masks to prevent any contamination of the film from 
lithographic processing. Contacts were made using Silver paste. Measurements 
were carried out in a Oxford 2K Helium cryostat with a base temperature of 
1.6 K and equipped with a 8 T superconducting magnet. Standard low frequency AC Lock-in technique was used for resistance measurements. Sample bias currents were in the 
range of 100 nA to 1 $\mu$A to avoid sample heating or high bias effects. 
The devices were aligned in different field configurations using specially 
designed sample holders for accurate sample placement. During in-plane 
magnetic field measurements, the samples were aligned such that the current 
direction was perpendicular to the direction of the in-plane magnetic field.

\vspace{0.5cm}
{\bf Acknowledgements}\\
K.M. thanks CSIR, India for financial support. D.S. thanks DST, India for support under Grant No. SR/S2/JCB-44/2010. P.S.A.K. acknowledges Nanomission, DST, Govt. of India for support.



\end{document}